\documentclass[runningheads]{llncs}
\usepackage[T1]{fontenc}
\usepackage{graphicx}
\usepackage[hidelinks]{hyperref}
\usepackage{amsmath}
\usepackage{booktabs}
\usepackage{multirow}
\usepackage{bbding}
\usepackage{orcidlink}
\usepackage{color}

\begin{document}
\title{Head and Neck Tumor Segmentation of MRI from Pre- and Mid-radiotherapy with Pre-training, Data Augmentation and Dual Flow UNet}
\titlerunning{Automatic Segmentation of H\&N Cancer for 3D MRI Image}
% If the paper title is too long for the running head, you can set
% an abbreviated paper title here
%
\author{Litingyu Wang\inst{1} \and
Wenjun Liao\inst{1,3} \and
Shichuan Zhang\inst{1,3} \and
Guotai Wang\inst{1,2(}\Envelope\inst{)}\orcidlink{0000-0002-8632-158X}}
\authorrunning{Wang et al.}
% First names are abbreviated in the running head.
% If there are more than two authors, 'et al.' is used.
%
\institute{University of Electronic Science and Technology of China, Chengdu, China \and
Shanghai AI Laboratory, Shanghai, China \and
Department of Radiation Oncology, Sichuan Cancer Hospital \& Institute, Sichuan Cancer Center, Chengdu, China\\
\email{guotai.wang@uestc.edu.cn}}
\maketitle              % typeset the header of the contribution
\begin{abstract}
Head and neck tumors and metastatic lymph nodes are crucial for treatment planning and prognostic analysis. Accurate segmentation and quantitative analysis of these structures require pixel-level annotation, making automated segmentation techniques essential for the diagnosis and treatment of head and neck cancer. In this study, we investigated the effects of multiple strategies on the segmentation of pre-radiotherapy (pre-RT) and mid-radiotherapy (mid-RT) images. For the segmentation of pre-RT images, we utilized: 1) a fully supervised learning approach, and 2) the same approach enhanced with pre-trained weights and the MixUp data augmentation technique. For mid-RT images, we introduced a novel computational-friendly network architecture that features separate encoders for mid-RT images and registered pre-RT images with their labels. The mid-RT encoder branch integrates information from pre-RT images and labels progressively during the forward propagation. We selected the highest-performing model from each fold and used their predictions to create an ensemble average for inference. In the final test, our models achieved a segmentation performance of 82.38\% for pre-RT and 72.53\% for mid-RT on aggregated Dice Similarity Coefficient (DSC) as HiLab. Our code is available at \url{https://github.com/WltyBY/HNTS-MRG2024_train_code}.

\end{abstract}
\section{Introduction}
Head and neck (H\&N) cancers are among the most prevalent types of cancer. Imaging in H\&N cancer serves multiple purposes, including quantitative assessment of tumors, evaluation of nodal disease, and differentiation between recurrent tumors and post-treatment changes~\cite{rumboldt2006imaging}. In recent years, convolutional neural networks (CNNs) within the realm of deep learning have profoundly impacted medical image analysis~\cite{badrigilan2021deep,chen2021transunet}. The detection and segmentation of H\&N tumors and metastatic lymph nodes are crucial for diagnosis and treatment. In addition, deep learning aids in these processes by segmenting target areas to facilitate treatment and surgical planning. During radiation treatment planning, computed tomography (CT) is commonly utilized~\cite{sager2019evaluation}. However, CT images often lack clear delineation between lymph nodes and surrounding tissues~\cite{braendengen2011delineation,dai2018state}. And, unfortunately, most existing datasets are based on CT, contrast-enhanced CT, or positron emission computed tomography (PET)~\cite{andrearczyk2021overview,luo2023segrap2023}. In contrast, the HNTS-MRG2024 Challenge provides a magnetic resonance imaging (MRI) dataset for H\&N images, representing an uncommon imaging modality in H\&N cancer diagnosis and capturing a significant scenario of adaptive radiotherapy. This competition focuses on segmenting gross tumor volumes of primary tumors (GTVp) and lymph nodes (GTVn) on both pre-radiotherapy (pre-RT) and mid-radiotherapy (mid-RT) MRI images, designated as Task-1 and Task-2, respectively.

Our methods and innovations are summarized as follows. We designed a novel framework, incorporating model pre-training, multiple data augmentations, and network architectures. Firstly, under the competition rules allowing the use of external public datasets, we pre-trained our model on a CT dataset due to the absence of an appropriate H\&N MRI dataset. Given the significant differences in image intensities between CT and MRI, we used histogram matching strategy to align the intensity distributions in the preprocessing stage and employed nonlinear intensity transformations in the model pre-training stage. Secondly, to address the scarcity of foreground voxels in our dataset, we employed the MixUp~\cite{zhang2017mixup} data augmentation technique. This approach not only expanded the training dataset, but also enhanced the model's generalizability. Furthermore, different from the conventional single encoder-decoder network structure commonly used in segmentation tasks, we adopted a dual encoder-decoder architecture to leverage per-RT images to guide the segmentation in mid-RT images. Attention blocks were introduced to this innovative structure, allowing for the effective fusion of information from disparate encoder streams. After conducting a rigorous five-fold cross-validation, we identified the top-performing model in each fold. The selected models demonstrated impressive performance, with an average aggregated Dice Similarity Coefficient (DSC) of 80.65\% for Task-1 and 74.68\% for Task-2 across five-fold cross-validation. Furthermore, in the final test, they scored 82.38\% for Task-1 and 72.53\% for Task-2.
\begin{figure}[t]
    \centering
    \includegraphics[width=0.9\textwidth]{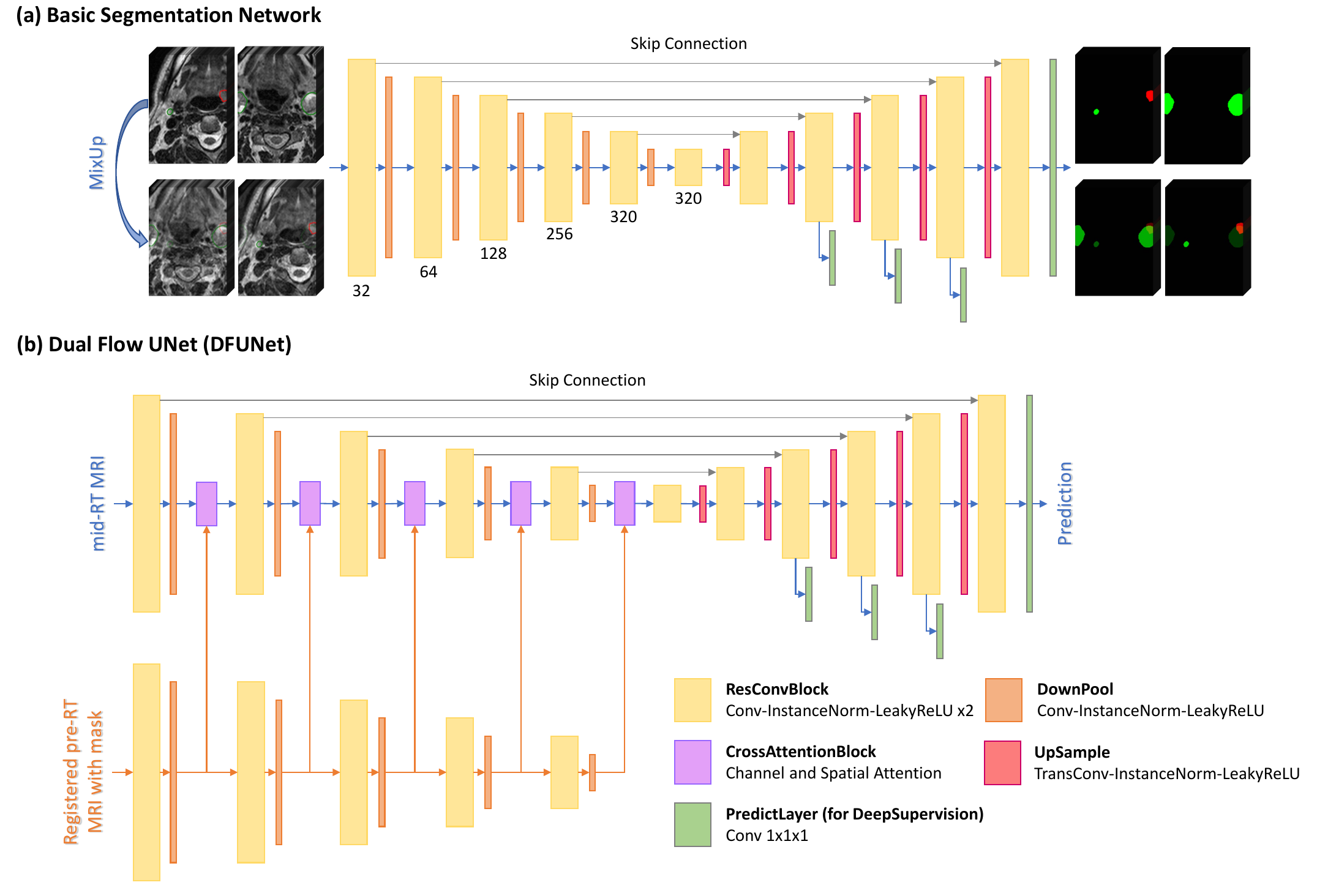}
    \caption{Two architectures used for training. (a) An encoder-decoder network, named basic segmentation network, with MixUp diagram. (b) Two separate encoders with one decoder named Dual Flow UNet (DFUNet).}
    \label{Network}
\end{figure}

\section{Methods}
We proposed multiple training strategies for the two tasks, and selected the best-performing model in each of the five-fold cross-validation divisions. For the segmentation of pre-RT images in Task-1, we employed:
\begin{itemize}
    \item[$\bullet$] Fully supervised learning utilizing an encoder-decoder architecture, as depicted in Fig.~\ref{Network}~(a).
    \item[$\bullet$] The same fully supervised learning approach, enhanced with pre-trained weights and the MixUp data augmentation method.
\end{itemize}
For Task-2, we introduced an additional training method:
\begin{itemize}
    \item[$\bullet$] Fully supervised learning employing the Dual Flow UNet (DFUNet) architecture, illustrated in Fig.~\ref{Network}~(b).
\end{itemize}

\subsection{Networks} 
\begin{figure}[t]
    \centering
    \includegraphics[width=0.7\textwidth]{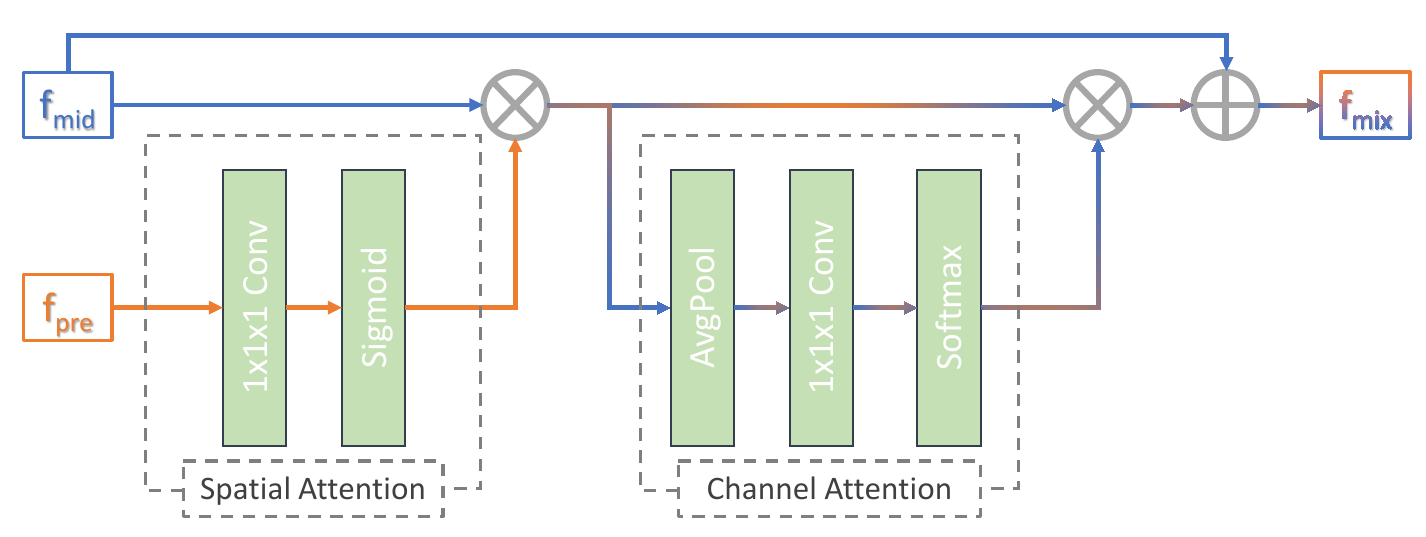}
    \caption{CNN-based cross attention block to integrate secondary information $f_{pre}$ into primary information $f_{mid}$.}
    \label{CrossAttention}
\end{figure}

We designed two network architectures in this study. The first network is basic segmentation network, shown as Fig.~\ref{Network}~(a)), which is an encoder-decoder architecture. For task-1, the input of basic segmentation network is the single-channel pre-RT images. For Task-2, the input of basic segmentation network is three-channel images, which are consisted of mid-RT image, registered pre-RT image and its mask. 
The second network is DFUNet, shown as Fig.~\ref{Network}~(b), which is only used for Task-2. The DFUNet comprises two encoders, whose inputs are a single-channel mid-RT image and a two-channel input (registered pre-RT image and its mask), respectively. To facilitate information fusion, the information from the pre-RT image and its mask is integrated into the mid-RT stream at various encoder stages using a CNN-based cross attention block~\cite{sun2022efficient}. This CNN-based attention mechanism differs from the attention blocks described in~\cite{vaswani2017attention} by employing both spatial and channel attention~\cite{woo2018cbam}. This approach enhances or fuses features while significantly reducing the computational demands of training and inference. The operational details of the cross attention block are illustrated in Fig.~\ref{CrossAttention}.
Details of the network's configuration are elaborated in Section~\ref{section:Model_details}.

\begin{figure}[t]
    \centering
    \includegraphics[width=0.7\textwidth]{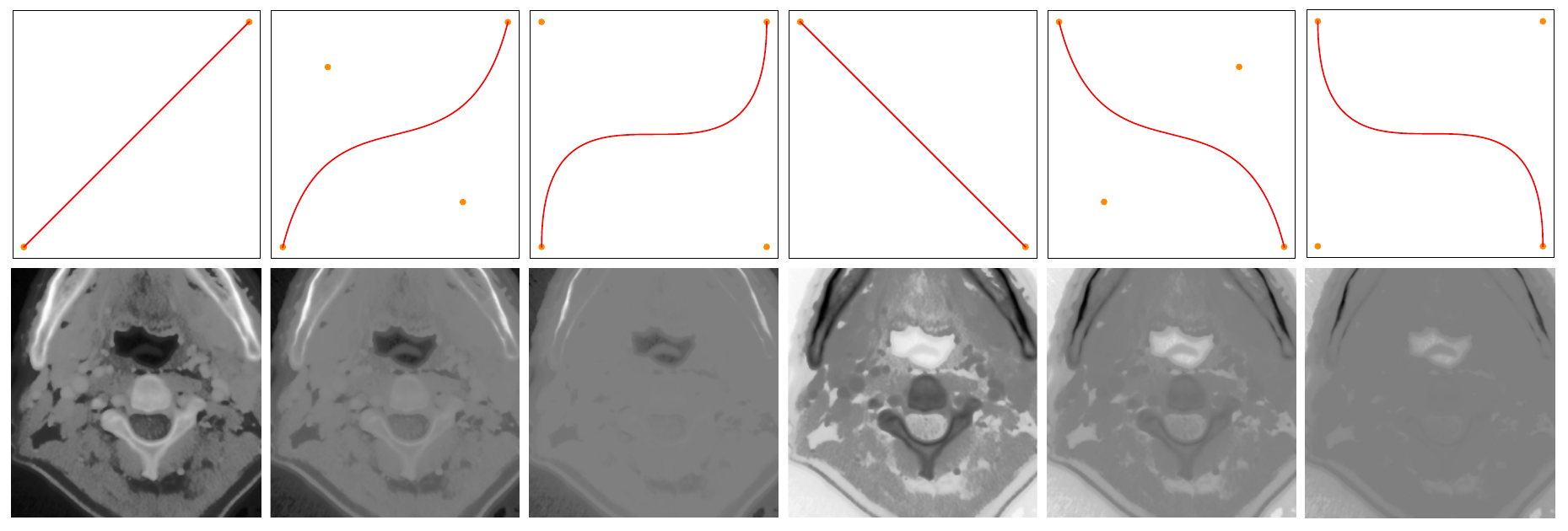}
    \caption{Raw image (the first column) and corresponding augmented images using B\'{e}zier Curve~\cite{liu2024rpl}.}
    \label{Nonlinear}
\end{figure}

\subsection{Model Pre-training}
We utilized the SegRap2023 Challenge Dataset~\cite{luo2023segrap2023} and the basic segmentation network depicted in Fig.\ref{Network}~(a) for fully supervised pre-training. However, since SegRap2023 is based on CT images, we encountered challenges due to the inherent intensity differences between CT and MRI data. To minimize these differences, we employed histogram matching during preprocessing to align the intensity distributions of CT and MRI as closely as possible. Despite this effort, intensity variations across various anatomical structures remained a challenge. To further address these issues, we incorporated techniques from domain generalization~\cite{zhou2022generalizable}, applying nonlinear transformations to the image intensities. This approach was designed to reduce the model's reliance on specific intensity values, thereby enhancing its generalization capabilities. The effectiveness of these nonlinear transformations is visualized in Fig.~\ref{Nonlinear}.

\subsection{MixUp Augmentation Strategy}
Since our training methodology is based on patches, patches in one mini-batch inevitably contain negative samples, which entirely belong to the background category. To mitigate the impact of these negative samples and to encourage the network to learn the distribution across different classes, we employ the MixUp~\cite{zhang2017mixup} technique to augment our dataset. MixUp creates new training examples by linearly interpolating between pairs of samples, which helps the model to generalize better across class boundaries. The process is defined by the following equations:
\begin{equation}
    \begin{split}
        \Tilde{x}=\lambda x_{i} + (1-\lambda) x_{j} \\
        \Tilde{y}=\lambda y_{i} + (1-\lambda) y_{j}
    \end{split}
\end{equation}
where $x_{i}$ and $x_{j}$ are raw input patches from different images, $y_{i}$ and $y_{j}$ are the corresponding one-hot labels, and $\lambda$ is a parameter in the range $[0,1]$ that controls the interpolation. In our experiments, $\lambda$ is sampled from the Beta distribution, which provides a flexible way to control the mixing ratio. During training, both the newly created examples from MixUp and the original samples are used in each batch, allowing the model to learn from a more diverse set of examples.

\subsection{Loss}
The inputs to our network can be categorized into two types: the raw input patches and the new cases generated by MixUp. For the raw input patches, we utilize a combination of CrossEntropy loss $\mathcal{L}_{CE}$ and Dice loss $\mathcal{L}_{Dice}$ for both pre-training and downstream training:
\begin{equation}
    Loss_{raw}=\sum^{l}_{d=0} \frac{1}{2^{d}} (\mathcal{L}_{CE}(x_{d}, y_{d}) + {L}_{Dice}(x_{d}, y_{d}))
\end{equation}
where $d$ represents the identifier for different resolutions, with smaller $d$ values corresponding to higher resolutions of $x_{d}$ and $y_{d}$. $l$ represents the number of resolutions used in computing the loss. The low-resolution labels $y_{d}$ are obtained by down-sampling the original labels. In contrast, for the cases generated through MixUp, we only compute $\mathcal{L}_{CE}$ and omit deep supervision:
\begin{equation}
    Loss_{MixUp}= \mathcal{L}_{CE}(\Tilde{x}, \Tilde{y})
\end{equation}
\begin{figure}[t]
    \centering
    \includegraphics[width=0.7\textwidth]{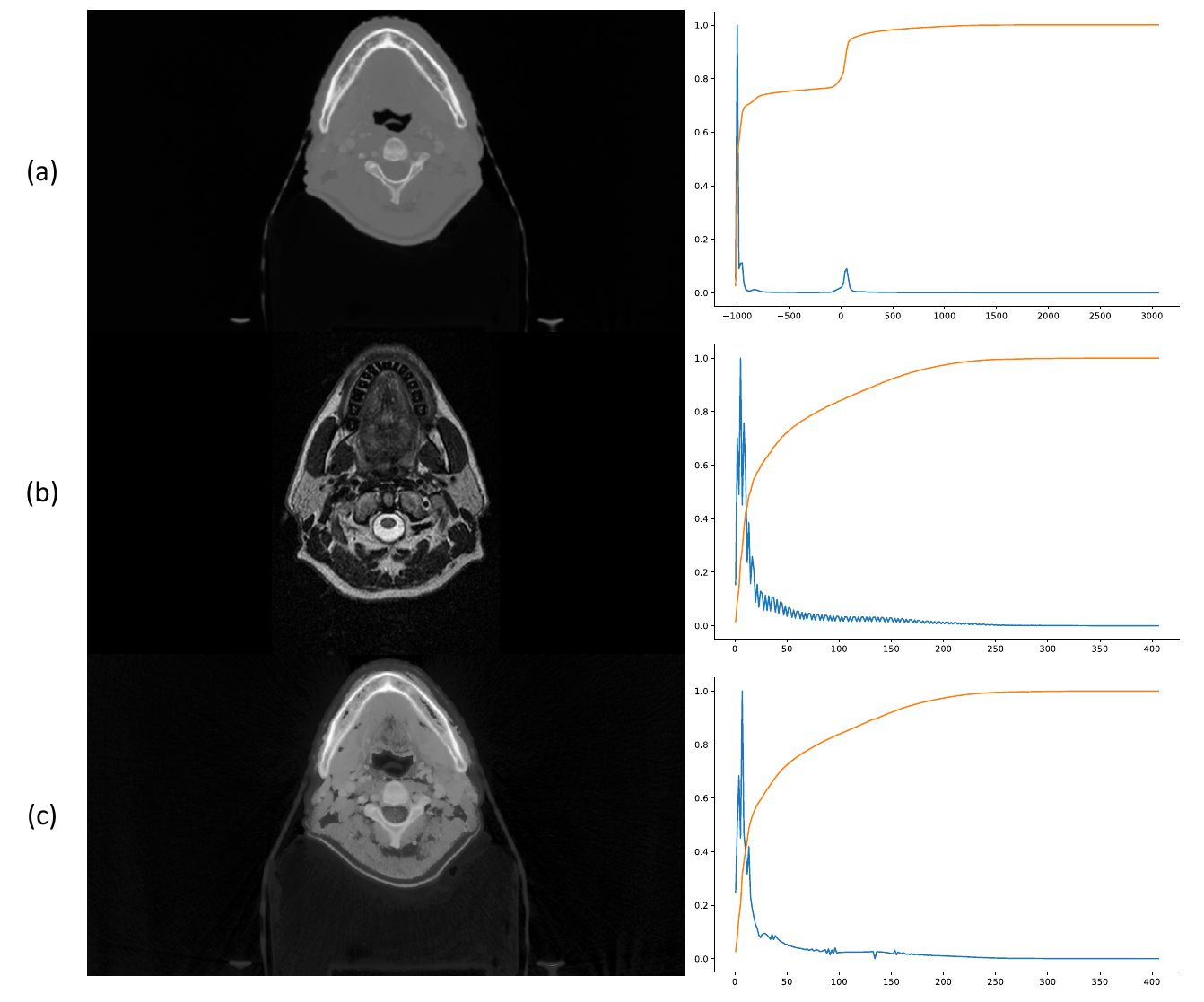}
    \caption{Histogram matching on the SegRap2023 dataset. (a) shows the source image and its grayscale histogram from the SegRap2023 dataset. (b) displays the reference image from the HNTS-MRG2024 dataset, and (c) presents the matched image.}
    \label{Matched_his}
\end{figure}

\section{Experiments}
We conducted our experiments using the nnUNetv2~\cite{isensee2021nnu} framework, which streamlined the hyperparameter selection process. This toolkit automates many decisions, allowing us to focus on the specific architectural adjustments needed for our tasks. As depicted in Fig.~\ref{Network}, we selected a VNet-like~\cite{milletari2016v} architecture for our network, which was trained using deep supervision.

\subsection{Datasets}
\subsubsection{SegRap2023 Challenge Dataset} Segmentation of Organs-at-Risk and Gross Tumor Volume of NPC for Radiotherapy Planning Challenge Dataset~\cite{luo2023segrap2023} comprises 120 pairs of H\&N CT and contrast-enhanced CT images. We ultimately used only the 120 enhanced CT scans because, during the training process, we found that the model's performance with enhanced CT was superior to that with CT or a combination of both.

\subsubsection{HNTS-MRG2024 Challenge Dataset} Head and Neck Tumor Segmentation for MR-Guided Applications 2024 Challenge Dataset includes three types of 3D T2-weighted (T2w) MRI scans: pre-RT, mid-RT, and registered pre-RT images of the head and neck. We randomly split the entire training set of 150 cases into 5 folds, training a separate model for each fold.

\subsection{Data Preprocessing}
\subsubsection{SegRap2023} First, we performed morphological operations to crop the images to the region of interest within the human body based on intensity. Second, we randomly selected a pre-RT image from the HNTS-MRG2024 dataset and applied histogram matching to align the grayscale histograms, as illustrated in Fig.~\ref{Matched_his}. Finally, we applied Z-score standardizing and resampled each volume into a resolution of $1.2mm \times 0.5mm \times 0.5mm$.

\subsubsection{HNTS-MRG2024} We selected pixels with intensity values greater than 60 to identify the region of interest. We then located the largest connected component and applied morphological operations to create a mask of the human body. Next, we cropped the image along the $x$ and $y$ axes to match the dimensions of the head area. Finally, we normalized the image using Z-score standardization and resampled it to the same resolution used for the SegRap2023 dataset.

\subsection{Models}
\label{section:Model_details}
As illustrated in Fig.~\ref{Network}, we proposed two distinct network architectures. Fig.~\ref{Network}~(a) displays a basic encoder-decoder model designed for semantic segmentation, which consists of six resolution stages. Each stage includes two convolutional layers with instance normalization, a LeakyReLU activation function, and a residual connection. This is followed by a DownPool block that contains an additional convolutional layer, instance normalization, and LeakyReLU activation. The structure of the decoder is symmetrical to that of the encoder, but it utilizes UpSample with transposed convolution. Features are pooled five times along the $x$ and $y$ axes, and three times along the $z$ axis. In Fig.~\ref{Network}~(b), the Dual Flow UNet (DFUNet) introduces an additional encoder and cross-attention blocks for merging two data streams. Both decoders share a design with four $1 \times 1 \times 1$ convolutions to output three channels for deep supervision.

\subsection{Implementations}
All the models were trained for 1000 epochs with a batch size of 2. When applying the MixUp~\cite{zhang2017mixup} augmentation strategy, the batch size was increased to 4, as illustrated in Fig.~\ref{Network}~(a). We used the Stochastic Gradient Descent (SGD) optimizer, with a momentum of 0.99, an initial learning rate of 0.01, and a weight decay of $3\times10^{-5}$. Training was conducted on a single NVIDIA RTX 4090 with 24GB of VRAM, using a patch size of $56\times224\times160$. Inference was performed with a sliding window strategy, enhanced by Test Time Augmentation (TTA) that included operations such as axes flips. Model performance was evaluated based on the aggregated Dice Similarity Coefficient (DSC).
\section{Results}
The results of the various training strategies based on a five-fold cross-validation of the training set, as outlined in the methods section, are detailed in Table~\ref{tab:Task-1} for Task-1 and Table~\ref{tab:Task-2} for Task-2. The bolding in Table~\ref{tab:Task-1} and Table~\ref{tab:Task-2} indicates the models selected for the corresponding data split, representing the models that yielded the best results in those particular folds. Overall, the average cross-validation performances for the selected models, as measured by the aggregated DSC, are 80.65\% for Task-1 and 74.68\% for Task-2.

\begin{table}[t]
    \centering
    \caption{The aggregated DSC (\%) values from ablation study of our method, based on 5-fold cross-validation for \textbf{Task-1}. The terms "Base", "Pre-train", "MixUp", and "Pre-train+MixUp" correspond to fully supervised training, fully supervised training with pre-trained weights, fully supervised training with MixUp augmentation, and a combination of pre-trained weights with MixUp, respectively.}
    \setlength{\tabcolsep}{2mm}{
        \begin{tabular}{c|cc|cc|cc|cc}
            \hline
            \multirow{2}[4]{*}{} & \multicolumn{2}{c|}{Base} & \multicolumn{2}{c|}{Pre-train} & \multicolumn{2}{c|}{MixUp} & \multicolumn{2}{c}{Pre-train+MixUp} \\
            \cline{2-9}          & GTVp  & GTVn  & GTVp  & GTVn  & GTVp  & GTVn  & GTVp  & GTVn \\
            \hline
            Fold1 & 78.63  & 83.89  & 78.60  & 83.44  & 81.64  & 84.81  & \textbf{82.07} & \textbf{84.33} \\
            Fold2 & 67.51  & 82.54  & 68.31  & 82.95  & 65.94  & 84.87  & \textbf{67.75} & \textbf{84.88} \\
            Fold3 & 70.68  & 84.86  & 71.10  & 86.55  & 70.21  & 85.79  & \textbf{70.65} & \textbf{87.19} \\
            Fold4 & 80.59  & 84.20  & 80.48  & 84.45  & 80.78  & 84.94  & \textbf{80.88} & \textbf{84.72} \\
            Fold5 & \textbf{80.39} & \textbf{83.60} & 77.66  & 84.70  & 79.80  & 84.02  & 78.82  & 84.48  \\
            \hline
            Average & 75.56  & 83.82  & 75.23  & 84.42  & 75.67  & 84.89  & \textbf{76.03}  & \textbf{85.12}  \\
            \hline
        \end{tabular}%
    }
    \label{tab:Task-1}%
\end{table}%

\begin{table}[t]
    \centering
    \caption{The aggregated DSC (\%) values of ablation study of our method, based on 5-fold cross-validation for \textbf{Task-2}. "Base" refers to fully supervised learning using mid-RT as inputs. "Base+pre-RT" signifies that the inputs have been expanded to include both mid-RT and registered pre-RT along with their labels. "DFUNet" indicates the substitution of the model with the DFUNet architecture compared to "Base+pre-RT".  "Pre-train+MixUp" means the experiments were conducted on the basic segmentation network depicted in Fig.\ref{Network}~(a) with pre-trained weights and MixUp augmentation.}
    \setlength{\tabcolsep}{2mm}{
        \begin{tabular}{c|cc|cc|cc|cc}
            \hline
            \multirow{2}[4]{*}{} & \multicolumn{2}{c|}{Base} & \multicolumn{2}{c|}{Base+pre-RT} & \multicolumn{2}{c|}{DFUNet} & \multicolumn{2}{c}{Pre-train+MixUp} \\
            \cline{2-9}          & GTVp  & GTVn  & GTVp  & GTVn  & GTVp  & GTVn  & GTVp  & GTVn \\
            \hline
            Fold1 & 40.46  & 74.10  & 59.80  & 86.62  & \textbf{64.48} & \textbf{86.82} & 62.08  & 87.63  \\
            Fold2 & 27.34  & 66.65  & \textbf{59.01} & \textbf{85.00} & 56.93  & 85.36  & 56.91  & 84.77  \\
            Fold3 & 38.21  & 75.47  & \textbf{65.46} & \textbf{86.96} & 58.74  & 86.65  & 64.98  & 87.45  \\
            Fold4 & 36.15  & 70.32  & 63.65  & 88.09  & 65.85  & 87.14  & \textbf{64.89} & \textbf{88.32} \\
            Fold5 & 45.85  & 79.72  & 58.66  & 86.59  & 55.87  & 86.92  & \textbf{58.24} & \textbf{87.55} \\
            \hline
            Average & 37.60  & 73.25  & 61.32  & 86.65  & 60.37  & 86.58  & \textbf{61.42}  & \textbf{87.14}  \\
            \hline
        \end{tabular}%
    }
    \label{tab:Task-2}%
\end{table}%

\begin{figure}[t]
    \centering
    \includegraphics[width=0.9\textwidth]{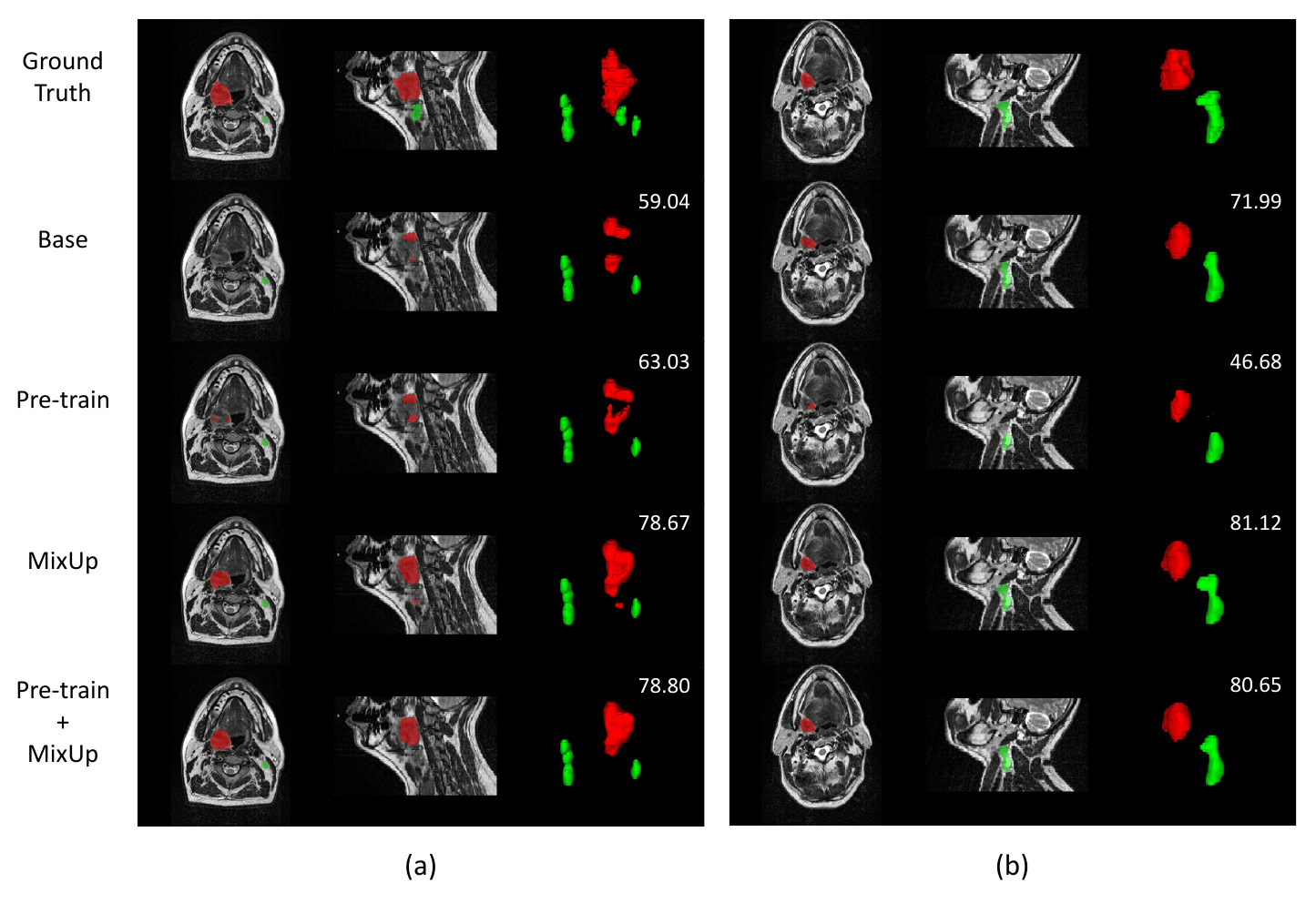}
    \caption{Examples of \textbf{Task-1} illustrate the segmentation of GTVp (red) and GTVn (green) with mean DSC values (numbers in white). (a) represents a well-predicted case, while (b) shows a poorly predicted one. Classification of samples as well- or poor-predicted here refers to whether they meet the method's improvements.}
    \label{vis_pre}
\end{figure}

For Task-1, as shown in Table~\ref{tab:Task-1}, all strategies struggle to achieve significant improvements in GTVp segmentation, although they positively impact GTVn. The combination of pre-trained weights and MixUp marginally enhance GTVp segmentation by 0.47\% in aggregated DSC, while GTVn sees a more substantial improvement of 1.30\%. As illustrated in Fig.\ref{vis_pre}(a), pre-trained weights facilitate goal exploration to some extent, whereas MixUp significantly aids in the identification of foreground categories.

\begin{figure}[t]
    \centering
    \includegraphics[width=0.9\textwidth]{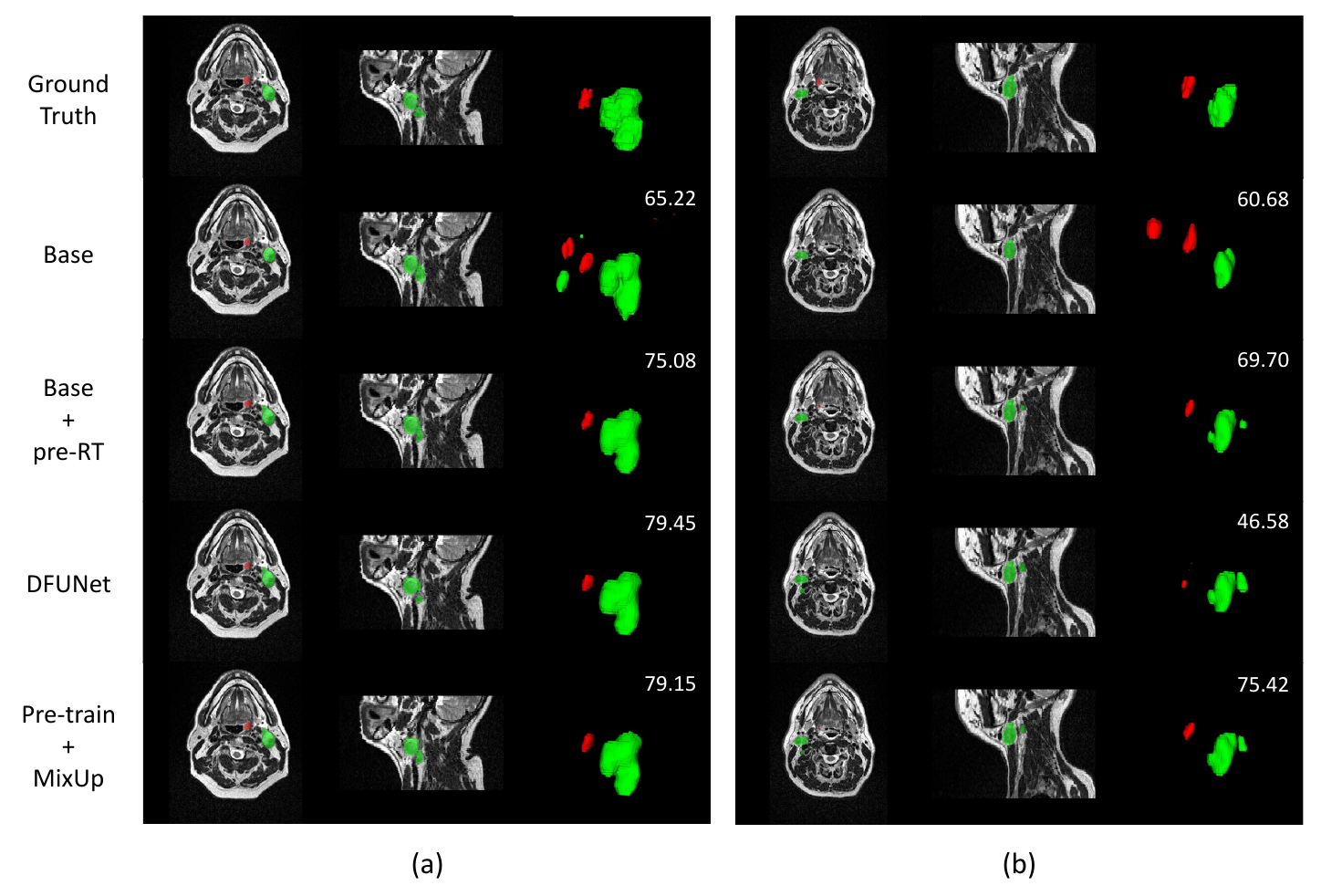}
    \caption{Examples of \textbf{Task-2} illustrate the segmentation of GTVp (red) and GTVn (green) with mean DSC values (numbers in white). (a) represents a well-predicted case, while (b) is the bad one.}
    \label{vis_mid}
\end{figure}

In Task-2, the situation is more complex. Although incorporating pre-RT images and their corresponding labels significantly enhances the segmentation performance, leading to an 18.56 percentage point increase in the average aggregated DSC, other methods show minimal improvement for GTVp segmentation, as demonstrated in Table~\ref{tab:Task-2}. The segmentation results for the case shown in Fig.\ref{vis_mid}(a) follow the design of our method. Notably, DFUNet outperforms the basic segmentation network in fold 1 and 4 of Table~\ref{tab:Task-2}, with average aggregated DSC improvements of 2.44\% and 0.63\%, respectively, which is not observed in other methods. In contrast, its performance is significantly worse in the other folds.

In the final testing phase, we selected the models that demonstrated the best compromise in performance and integrated their predictions, which are highlighted in Table~\ref{tab:Task-1} and Table~\ref{tab:Task-2}. The final test allowed for a single submission, with our results for both tasks summarized in Table~\ref{tab:Final Test}.

\begin{table}[h]
    \centering
    \caption{Results of Final Test for the two tasks on aggregated DSC (\%)}
    \setlength{\tabcolsep}{2mm}{
        \begin{tabular}{c|c|c|c}
        \hline
              & GTVp  & GTVn  & Average \\
        \hline
        Task-1 & 78.51 & 86.24 & 82.38 \\
        Task-2 & 57.95 & 87.12 & 72.53 \\
        \hline
        \end{tabular}%
    }
    \label{tab:Final Test}%
\end{table}%

\section{Discussion and Conclusion}
In this study, we developed GTV segmentation models using two variations of the VNet-like~\cite{milletari2016v} architecture on the MRI T2w H\&N dataset from the HNTS-MRG2024 Challenge. Our investigation focused on expanding the dataset and refining network structures. To augment the dataset, we initially used an external public CT dataset~\cite{luo2023segrap2023} for pre-training and subsequently applied MixUp technology to generate more training data. For Task-2, we also utilized additional registered pre-RT images and masks to enhance the segmentation of mid-RT. Beyond the standard segmentation network structure, we introduced the DFUNet, which includes two encoders, one decoder, and CNN-based cross attention blocks. This architecture enables the network to differentiate between the primary image for segmentation and supplementary information.

The 5-fold cross-validation results indicate that our proposed strategies have led to significant improvements in the segmentation of simpler categories, such as GTVn. However, the advancements in the segmentation of a more challenging category, like GTVp, have been marginal and could even introduce adverse effects, especially for Task-2, as shown in Fig.~\ref{vis_mid}~(b). We suspect that the suboptimal learning of GTVp is due to the severe class imbalance and the learning of full background samples. While nnUNetv2~\cite{isensee2021nnu} samples foreground categories during training, its methodology does not adequately address the imbalance among foreground categories. A potential solution might involve an expanded oversampling strategy that targets each foreground and background category individually. Although the DFUNet underperformed compared to the basic segmentation model in cross-validation, it demonstrated notable enhancements in GTVp segmentation in some folds. We are confident that with further refinements to the DFUNet architecture, it could outperform the basic model. Additionally, it is worth noting that pre-training is not universally effective, and in some cases, the use of pre-trained weights can significantly degrade segmentation performance. As illustrated in Fig.~\ref{vis_pre}~(b), the weaker segmentation performance of GTVn in CT scans may hinder the model's ability to segment MRI samples that are difficult to segment for GTVn.

Furthermore, during pre-training with the SegRap2023 Challenge dataset~\cite{luo2023segrap2023}, we observed that the segmentation performance for GTVp was markedly higher on the CT dataset compared to GTVn. Interestingly, this trend was reversed for T2-weighted MRI in HNTS-MRG2024. Although the types of cancer in the two datasets are different, a model that can combine the advantages of the two modalities can improve the segmentation performance while greatly reducing the dependence on multi-modal data and minimizing the patient's exposure during image acquisition. Potential approaches to achieve this include cross-modal distillation~\cite{wang2023learnable}, domain adaptation~\cite{liu2021automated}, or the use of generative models~\cite{xia2024robust}.

In conclusion, we investigated the enhancement of fully supervised learning through dataset expansion and domain generalization techniques. We observed a notable improvement in the performance of GTVn with pre-training and the application of the MixUp strategy. Additionally, DFUNet led to significant enhancements in GTVp segmentation in some folds. Our average aggregated DSC across the folds is 80.65\% for Task-1 and 74.68\% for Task-2. In the final test, we achieve scores of 82.38\% for Task-1 and 72.53\% for Task-2. Future work could concentrate on refining the DFUNet architecture, effectively initializing its weights, and leveraging pre-RT data to boost the segmentation performance of mid-RT.

\begin{credits}
\subsubsection{\ackname} This work was supported by the Radiation Oncology Key Laboratory of Sichuan Province Open Fund (2022ROKF04).

\subsubsection{\discintname}
The authors have no competing interests to declare that are relevant to the content of this article.
\end{credits}

%
% ---- Bibliography ----
%
% BibTeX users should specify bibliography style 'splncs04'.
% References will then be sorted and formatted in the correct style.
%
% \bibliographystyle{splncs04}
% \bibliography{mybibliography}
%
% \begin{thebibliography}{8}

\nocite{*}
\bibliography{Reference}

\begin{thebibliography}{10}
\providecommand{\url}[1]{\texttt{#1}}
\providecommand{\urlprefix}{URL }
\providecommand{\doi}[1]{https://doi.org/#1}

\bibitem{andrearczyk2021overview}
Andrearczyk, V., Oreiller, V., Boughdad, S., Rest, C.C.L., Elhalawani, H., Jreige, M., Prior, J.O., Valli{\`e}res, M., Visvikis, D., Hatt, M., et~al.: Overview of the hecktor challenge at miccai 2021: automatic head and neck tumor segmentation and outcome prediction in pet/ct images. In: 3D head and neck tumor segmentation in PET/CT challenge, pp. 1--37. Springer (2021)

\bibitem{badrigilan2021deep}
Badrigilan, S., Nabavi, S., Abin, A.A., Rostampour, N., Abedi, I., Shirvani, A., Ebrahimi~Moghaddam, M.: Deep learning approaches for automated classification and segmentation of head and neck cancers and brain tumors in magnetic resonance images: a meta-analysis study. International journal of computer assisted radiology and surgery  \textbf{16},  529--542 (2021)

\bibitem{braendengen2011delineation}
Br{\ae}ndengen, M., Hansson, K., Radu, C., Siegbahn, A., Jacobsson, H., Glimelius, B.: Delineation of gross tumor volume (gtv) for radiation treatment planning of locally advanced rectal cancer using information from mri or fdg-pet/ct: a prospective study. International Journal of Radiation Oncology* Biology* Physics  \textbf{81}(4),  e439--e445 (2011)

\bibitem{chen2021transunet}
Chen, J., Lu, Y., Yu, Q., Luo, X., Adeli, E., Wang, Y., Lu, L., Yuille, A.L., Zhou, Y.: Transunet: Transformers make strong encoders for medical image segmentation. arXiv preprint arXiv:2102.04306  (2021)

\bibitem{dai2018state}
Dai, Y., King, A.: State of the art mri in head and neck cancer. Clinical radiology  \textbf{73}(1),  45--59 (2018)

\bibitem{isensee2021nnu}
Isensee, F., Jaeger, P.F., Kohl, S.A., Petersen, J., Maier-Hein, K.H.: nnu-net: a self-configuring method for deep learning-based biomedical image segmentation. Nature methods  \textbf{18}(2),  203--211 (2021)

\bibitem{liu2021automated}
Liu, J., Liu, H., Gong, S., Tang, Z., Xie, Y., Yin, H., Niyoyita, J.P.: Automated cardiac segmentation of cross-modal medical images using unsupervised multi-domain adaptation and spatial neural attention structure. Medical Image Analysis  \textbf{72},  102135 (2021)

\bibitem{liu2024rpl}
Liu, X., Wu, J., Luo, X., Liao, W., Zhang, S., Zhang, S., Wang, G.: Rpl-sfda: Reliable pseudo label-guided source-free cross-modality adaptation for npc gtv segmentation. In: 2024 IEEE International Symposium on Biomedical Imaging (ISBI). pp.~1--5. IEEE (2024)

\bibitem{luo2023segrap2023}
Luo, X., Fu, J., Zhong, Y., Liu, S., Han, B., Astaraki, M., Bendazzoli, S., Toma-Dasu, I., Ye, Y., Chen, Z., et~al.: Segrap2023: A benchmark of organs-at-risk and gross tumor volume segmentation for radiotherapy planning of nasopharyngeal carcinoma. arXiv preprint arXiv:2312.09576  (2023)

\bibitem{milletari2016v}
Milletari, F., Navab, N., Ahmadi, S.A.: V-net: Fully convolutional neural networks for volumetric medical image segmentation. In: 2016 fourth international conference on 3D vision (3DV). pp. 565--571. Ieee (2016)

\bibitem{rumboldt2006imaging}
Rumboldt, Z., Gordon, L., Bonsall, R., Ackermann, S.: Imaging in head and neck cancer. Current treatment options in oncology  \textbf{7},  23--34 (2006)

\bibitem{sager2019evaluation}
Sager, O., Dincoglan, F., Demiral, S., Gamsiz, H., Uysal, B., Ozcan, F., Colak, O., Dirican, B., Beyzadeoglu, M.: Evaluation of the impact of magnetic resonance imaging (mri) on gross tumor volume (gtv) definition for radiation treatment planning (rtp) of inoperable high grade gliomas (hggs). Concepts in Magnetic Resonance Part A  \textbf{2019}(1),  4282754 (2019)

\bibitem{sun2022efficient}
Sun, J., Dai, Y., Zhang, X., Xu, J., Ai, R., Gu, W., Chen, X.: Efficient spatial-temporal information fusion for lidar-based 3d moving object segmentation. In: 2022 IEEE/RSJ International Conference on Intelligent Robots and Systems (IROS). pp. 11456--11463. IEEE (2022)

\bibitem{vaswani2017attention}
Vaswani, A.: Attention is all you need. Advances in Neural Information Processing Systems  (2017)

\bibitem{wang2023learnable}
Wang, H., Ma, C., Zhang, J., Zhang, Y., Avery, J., Hull, L., Carneiro, G.: Learnable cross-modal knowledge distillation for multi-modal learning with missing modality. In: International Conference on Medical Image Computing and Computer-Assisted Intervention. pp. 216--226. Springer (2023)

\bibitem{woo2018cbam}
Woo, S., Park, J., Lee, J.Y., Kweon, I.S.: Cbam: Convolutional block attention module. In: Proceedings of the European conference on computer vision (ECCV). pp. 3--19 (2018)

\bibitem{xia2024robust}
Xia, Y., Feng, S., Zhao, J., Yuan, Z.: Robust cross-modal medical image translation via diffusion model and knowledge distillation. In: 2024 International Joint Conference on Neural Networks (IJCNN). pp.~1--8. IEEE (2024)

\bibitem{zhang2017mixup}
Zhang, H.: mixup: Beyond empirical risk minimization. arXiv preprint arXiv:1710.09412  (2017)

\bibitem{zhou2022generalizable}
Zhou, Z., Qi, L., Yang, X., Ni, D., Shi, Y.: Generalizable cross-modality medical image segmentation via style augmentation and dual normalization. In: Proceedings of the IEEE/CVF conference on computer vision and pattern recognition. pp. 20856--20865 (2022)

\end{thebibliography}
\bibliographystyle{splncs04}

% \end{thebibliography}
\end{document}